\documentclass[twocolumn]{aa}
\usepackage{graphicx}

\begin{document}

\title{Alignment of radio emission and quasars across NGC 613 and NGC 936 
and radio ejection from NGC 941 }

\author{H. Arp}

\offprints {H. Arp}

\institute{Max-Planck-Institut f\"ur Astrophysik, Karl Schwarzschild-Str.1,
  Postfach 1317, D-85741 Garching, Germany\\
   \email{arp@mpa-garching.mpg.de}}

\date{Received}

\abstract{NGC 613 (AM 0132-294) is a multi armed spiral with an active nucleus. 
Radio emission is aligned along the narrow inner bar of the spiral and appears to 
include inner and outer pairs of radio sources which define a line of ejection from 
this nucleus. Four high redshift quasars from the 2dF survey fall further out along 
this same line. The redshifts of these four, plus two others in this general 
direction, are close to the peak values in the periodic redshift distribution of 
quasars. 

In separation, redshifts and other properties there is an almost exact duplicate of
NGC 613 in the pair of $z \sim 2$ quasars paired across the radio ejecting, barrred 
galaxy NGC 936. 

NGC 941, a companion to NGC 936, has a quasar apparently 
connected to it by a radio bridge which shows structure at high resolution.

One of the pair across NGC 936, a strong Parks quasar, is also accurately aligned 
with a strong Parks quasar further away. Possible relationships between other 
quasars and low redshift galaxies in a broader region around NGC 936 are discussed.

}

\titlerunning{QSO-galaxy relations}

\maketitle

\section{Introduction}

In the Catalogue of Southern Peculiar Galaxies and Associations  (Arp \& Madore 
1987) the galaxy NGC 613 appears as a striking example of the class of multi-armed 
spirals (Fig. 1). NGC 613 (AM0132-294) is bright ($B_T= 10.75$), barred 
(SBb(rs)II) and has a redshift of z = .005. High resolution images show a narrow, 
high brightness bar extending from the nucleus out on either side, roughly NW 
and SE. 

The strong, flocculent spiral arms drew attention from some of the earliest 
investigators of internal motions. Spectra by E.M. Burbidge et al. (1964) revealed 
assymmetric velocities ranging up to 250 km/sec with respect to the nucleus. 
Later spectra by Hummel et al. (1987) indicated outflow in the nuclear emission 
lines. But already in 1964 E.M. Burbidge et al. had remarked that "For such a 
galaxy, the estimate of the total mass is highly uncertain."

Radio maps established the galaxy as a strong radio source (PKS 0131-296) and 
later Very Large Array, VLA maps, catalogued radio sources within the inner 
regions of the galaxy. It is also an infrared, IRAS source, and was listed as 
morphologically peculiar (Vorontsov-Velyaminov 824).

\begin{figure}
\includegraphics[width=8.6cm]{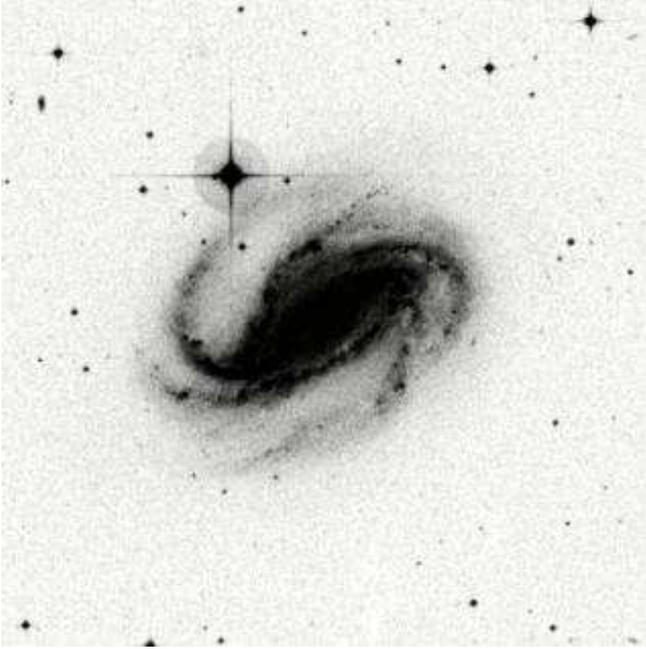}
\caption{UK Schmidt photograph of NGC 613, classified as a multi armed spiral, 
AM 0132-294. The frame is 9 x 9 arcmin.
\label{fig1}}
\end{figure}


\begin{figure}
\includegraphics[width=6.0cm]{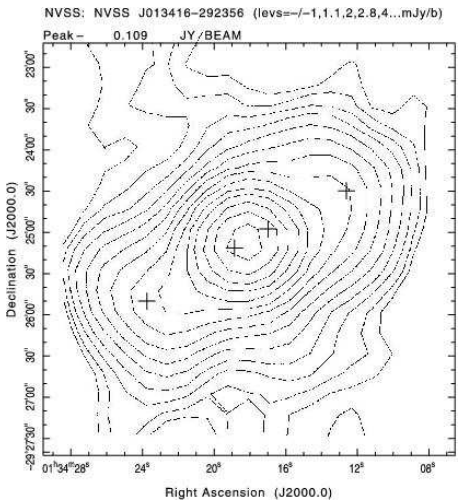}
\caption{VLA map of central 5 x 5 arcmin of NGC 613. The plus signs indicate the 
four catalogued NVSS radio sources. Distances and position angles of these 
sources are given in Table 1. The frame is 5 x 5 arcmin.
\label{fig2}}
\end{figure}

\begin{figure}[h]
\includegraphics[width=8.5cm]{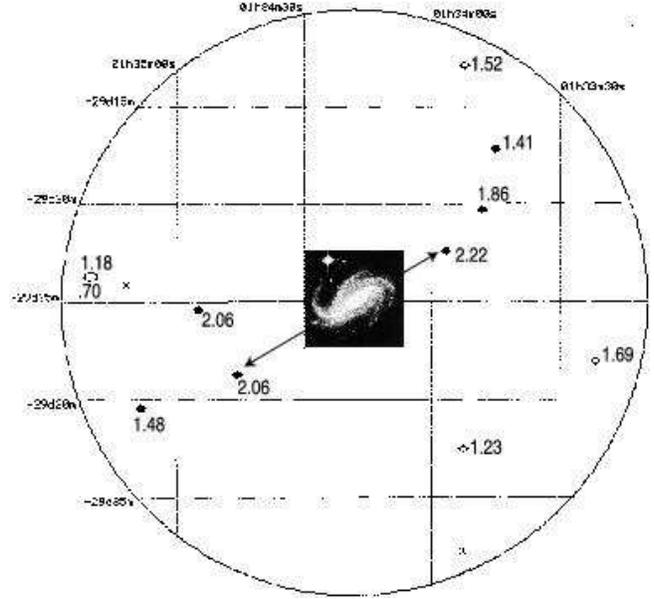}
\caption{Simbad map inside 15 arcmin radius around NGC 613. Quasars from the 
2dF Survey are labeled with their redshifts. Arrows indicate the direction of the 
radio ejections pictured in Fig. 2.
\label{fig3}}
\end{figure}

\section{Ejection of Radio Sources?}
The VLA maps at 1.49 MHz (Condon et al. 1998) identified four radio sources 
within NGC 613 whose distances from the center of NGC 613 are listed below. 
The contour plot with a beam size of $45^"$, however,  does not show 
conspicuous point sources (Fig.2).  The extraction software has succeeded 
in modeling most of the flux into point sources but there is the possibility that 
they may be minor fluctuations in extended emission and that the contour map 
does not represent true, compact point sources. 

On the other hand maps with $2^"$ resolution (Hummel et al 1987; Hummel and 
Jors\"ater 1992) show the radio emission in the interior $25^"$ breaking up into 
various concentrations and sources and it seems unlikely that the emission 
extending out past radii of $1.4^"$ is perfectly smooth. It would be important to 
confirm the existence of compact radio sources out at these distances 
situated along the narrow bar of the spiral. In the discussion which follows we 
will assume these NVSS catalogued positions represent sources of some degree 
compactness. In any case, the presence of this radio emission along the 
position angle of the bar strongly suggests its ejection from the innermost 
nucleus. 

\begin{table} \caption{Radio sources across NGC 613}
\label{Table1} \vspace{0.3cm}
\begin{tabular}{ccc}
Source & Dist.arcmin & p.a. (deg)\\
& & \\
NVSS J013418-292513 & 0.2 & 127\\
NVSS J013416-292456 & 0.3 & 298\\
NVSS J013412-292430 & 1.4 & 296\\
NVSS J013423-292551 & 1.4 & 123 \\
\end{tabular}
\end{table}

\begin{table} \caption{Quasars along radio ejection lines}
\label{Table 2} \vspace{0.3cm}
\begin{tabular}{cccc}
Quasar & Dist.arcmin & p.a. (deg) & z\\
& & &\\
2QZ J013356.8-292223 & 5.4 & 299 & 2.222\\
2QZ J013445.8-292842 & 7.0 & 121 & 2.059\\
2QZ J013454.8-292523 & 8.0 &  94 & 2.062\\
2QZ J0I3448.0- 292015 & 8.2 & 304 & 1.855 \\

2QZ J013345.0-291608 & 10.8 & 317 & 1.413\\
2QZ J013508.4-293023 & 12.2 & 117 & 1.482\\
\end{tabular}
\end{table}

\section{Quasars Along the Radio Ejection Lines}

Tables 1 and 2 show the distances and alignments of the radio sources and QSO's 
as measured from the center of NGC 613.
Fig. 3 shows that the nearest quasar to NGC 613 (z = 2.222) at $p.a. = 299^o$.
is accurately along the radio line. The next nearest quasar (z = 2.059) is at at 
$p.a. = 121^o$.

     It is unlikely to find a quasar so close to NGC 613 and also so closely along the line of 
radio ejection at $p.a.= 298^o$. But given that, it is even more unlikely to find the next 
closest quasar lying at $p.a = 121^o$, closely along the line of the opposite radio ejection.

      As for the quasars further out, they also tend to be placed along the ejection direction, 
particularly the z = 1.86 and z = 1.48 quasars. So there are two sets of quasars suggesting 
a match with the two inner pairs of radio sources.

     The numerical value of the quasar redshifts also argue against their being accidental 
background projections. We note particularly the the average redshift of the nearest two 
on the NW as being z = 2.04 closely matching the z = 2.06 of the ones on the SE. The two 
in the outer most pair are also very similar at z = 1.41 and z = 1.48. 

\begin{figure}
\includegraphics[width=8.5cm]{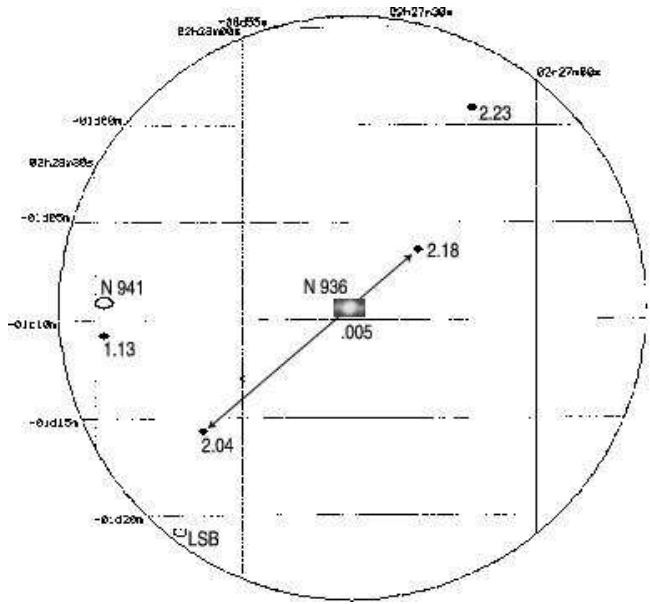}
\caption{Simbad map inside 15 arcmin radius around NGC 936. Quasars from the 
2dF Survey are labeled with their redshifts. Arrows indicate the direction of the 
radio ejections pictured in Fig. 5.
\label{fig4}}
\end{figure}

\section{Pairing across NGC 936}

An amazing coincidence is shown in Fig. 4 where there is an an almost identical 
configuration to that just discusssed in NGC 613. Even the redshifts of the 
quasars and the galaxies are very closely the same! Fig. 4 shows the SE quasar 
is z = 2.04 in NGC 936 compared to z = 2.06 for NGC 613. The NW quasar is 
z = 2.18 compared to z = 2.22 for NGC 613. Both central galaxies have z = .005 
and the separation of the pair across NGC 936 is 14.5 arcmin compared to 12.5 
arc min across NGC 613.

\begin{figure}
\includegraphics[width=8.5cm]{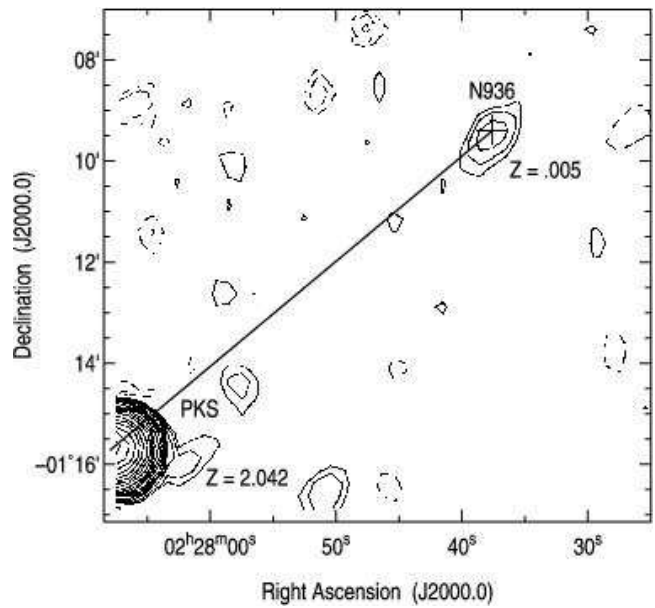}
\caption{The nvss map showing radio extension from NGC 936 along the line of 
adjacent quasars. The extension is stronger in the SE direction toward the PKS 
quasar of z = 2.04.
\label{fig5}}
\end{figure}

\subsection{Is NGC 936 an active galaxy?}

Fig. 5 shows an NVSS map of NGC 936. It is apparent that the galaxy has radio
isophotes elongated along the direction of the closeby pair of $z \sim 2$ quasars.
The elongation is greater in the direction of the z = 2.04 quasar which is a very
strong PKS radio source seen at the SE edge of Fig. 5. The only X-ray images of
NGC 936 are from the Einstein IPC. They show jet-like X-ray emission extending 
to the SE of the galaxy at a somewhat lesser position angle than the radio
extension. {\it Overall it seems reasonable to consider NGC 936 an active galaxy
showing signs of ejecting along the line of the pair, most strongly toward the strong 
PKS quasar at z = 2.04.}

As the insert to Fig. 4 shows, NGC 936 is a galaxy that is classified SB0 
(Sandage and Tamman 1981).  The important aspect, however seems to be the 
two large clumps extending on either side of a slightly irregular shaped 
nucleus, giving a shape of a fat bar.

\subsection{NGC 941, the companion to NGC 936}

Fig. 4 shows NGC 941, an ScdIII companion galaxy to NGC 936 at 204 km/sec 
greater redshift. A quasar about 2 arcmin S of it was found by Arp (1981) in a 
systematic search for UV excess objects near companion galaxies. (The working 
hypothesis at that time was that companion galaxies were more active than parents. 
This may be true since there was a very strong association of quasars with companion 
galaxies found - but in the ensuing years it has become clear that parent galaxies also 
can be very active.)

In any case NGC 941 is a radio source and Fig. 6 shows it has a strong plume of
radio material reaching down and enclosing the quasar. The quasar appears offset 
by 30 arcsec from the center of the nvss radio lobe and in order to interpret this 
configuration it is fortunate that there is a high resolution, FIRST map this region.

The FIRST extract shows that quasar at the center of the radio lobe is itself a radio 
source, slightly resolved. The stronger source about 30 arcsec WNW of the quasar is 
also compact but definitely extended toward the quasar. This would suggest that the 
stronger source (not optically identified to the limit of the dss plates) had been 
ejected from the quasar. That the quasar is ejecting radio material is supported by 
Fig. 7 which is a contrast enhanced FIRST extract which shows a small pair of radio 
sources aligned across the quasar image. (Plus some fainter possible sources.)

These observations suggest that we may be seeing details of secondary ejections 
involved in the evolution of quasars. It would therefore be an important group of 
radio sources to investigate with the highest resolution to the faintest isophote 
levels. As it stands at present, however, it seems likely that the z = 1.130 quasar is 
physically involved with ejection of radio material from NGC 941.

\begin{figure}
\includegraphics[width=8.5cm]{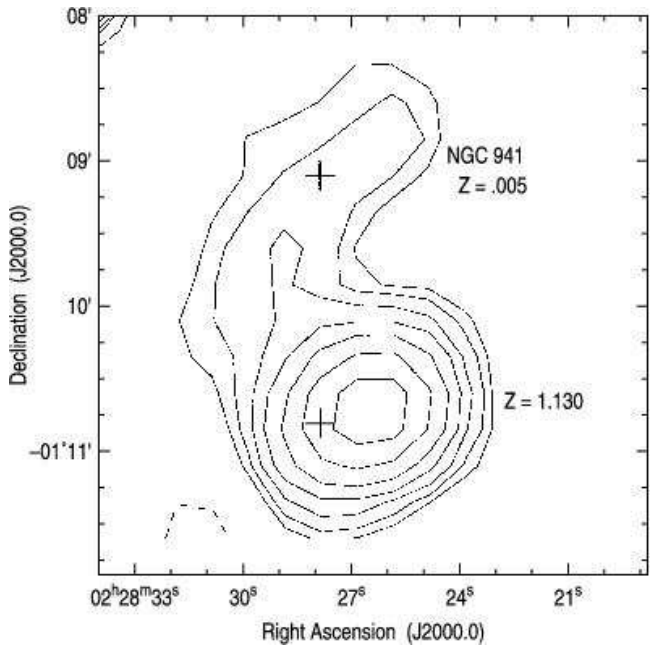}
\caption{This nvss map shows radio extension from NGC 941 down to, and 
enveloping the z = 1.130 quasar about 2 arcmin to the south.
\label{fig6}}
\end{figure}

\begin{figure}
\includegraphics[width=8.5cm]{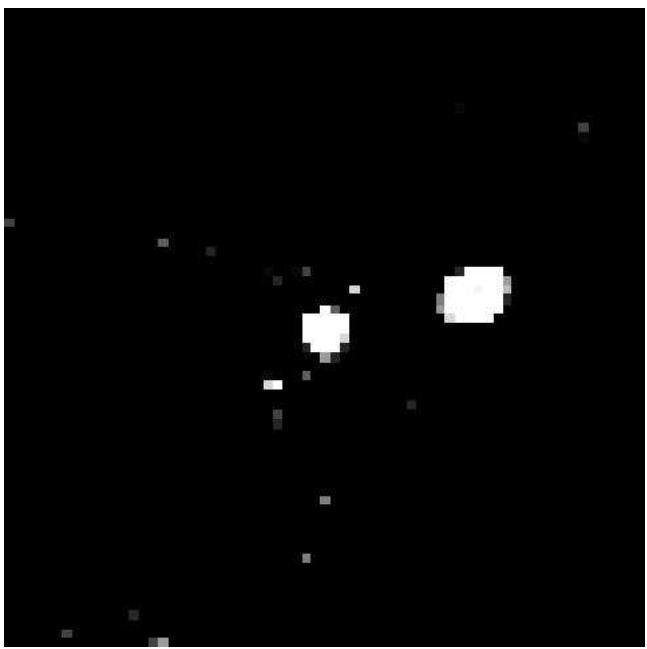}
\caption{This high resolution, FIRST map has been contrast enhanced to show the extension of
the brightest radio component back to the quasar and the small radio sources on either side of
the quasar.
\label{fig7}}
\end{figure}

\section{Relation to Previous Associations of Quasars with Low Redshift Galaxies}
     The pattern of quasars of similar redshift paired across active and ejecting galaxies of much 
lower redshift has been recorded many times now (Arp 1967; E. M. Burbidge 1997; Radecke 1997; 
Arp 1998; 2003; Arp et al. 2001; 2002). Quasars also have been associated with radio and X-ray 
ejections from the beginning. In the present case of NGC 613, however, the alignment 
of radio emission coinciding so closely with the alignment of the pairs of quasars is perhaps the
most striking evidence yet that quasars, like radio sources, are ejected from galaxy nuclei.

      In order to emphasize the significance of the patterns which are repeated in the association 
of the above quasars with NGC 613 and NGC 936, we outline a possible scenario as follows:

      {\bf 1}) Ejection of high intrinsic redshift plasmoids from the nucleus. 

      {\bf 2}) The radio emitting plasma is less dense and is stripped from the quasar as it travels 
outward.

This could result from a radio burst in the quasar while moving through the galactic medium or by 
encountering a cloud as it exits into the intergalactic medium.  Examples of this process have been 
discussed previously (Arp 1999a; 2001). It would furnish a physical model for why the quasar and radio 
sources are coincident and/or aligned in the early stages. Since the low density radio cloud would 
subsequently move under different forces, this model would also account for the paucity, in general, 
of radio loud quasars compared to X-ray and optical quasars.

It was concluded in the above references that the ejection of quasars or proto quasars through the 
disk disrupts the ejecting galaxy and slows their ejection velocities.  In the case of NGC 613 this 
would explain both the  non-equilibrium disturbances in the disk and why the quasars are found so 
unusually close the galaxy of origin.  (It is striking that the six quasars fall within a radius of from 5 to 
12 arcmin from NGC 613 whereas in most cases of association with large low redshift galaxies their 
quasars are distributed out to radii of 50 arcmin.)

     {\bf 3}) The redshifts decrease going away from nucleus and are quantized at the Karlsson 
(1971; 1990) peaks .

The redshifts of the six quasars fall from z = 2.22 to 1.41 as they go away
from the center. This decreasing redshift with separation has been observed in the best defined 
lines of quasars coming from the Seyfert galaxies NGC 3516 and NGC 5985 (Arp 1999b). The obvious
interpretation has been that the most recently ejected quasars are the youngest and therefore have
the highest redshift (Narlikar and Arp 1993; Arp 1998). But the NGC 613 quasars are obviously
quantized in redshift also. Two of the most populated Karlsson peak redshifts for quasars in general 
are z = 1.96 and 1.41. The four innermost quasars average to z = 2.05 and the two outer ones  
to z = 1.45. For some regions of the sky, for example the 12h, NGH region where the peaks are  at 
z = 2.05 and z = 1.48 (Arp et al. 1990), an even closer match results. 

     {\bf 4}) The redshifts tend to be numerically equal at equal distances on either side of the galaxy.

It has been observed previously that there is a strong tendency for quasars to occur in equi-distant, 
aligned pairs with the two quasars resembling each other closely in redshift. We see that again in the 
NGC 613 case, but exceptionally there are now two aligned pairs. The first pair consist of quasars of 
z = 2.22 and z = 2.06. The second pair consists of z = 1.48 and z = 1.41.  It is interesting to note that 
close to the z = 2.059 quasar lies a similar quasar of z = 2.062 - an unlikely coincidence if at 
cosmological distances. Next to the z = 2.22 quasar lies the z = 1.86 quasar. In both these cases it 
might be proposed that the original quasar had split, and one pair is separating across the line of sight 
and the other with a component of velocity along the line of sight.                       

     {\bf 5}) NGC 613 and spiral arms. 

The most unusual morphological feature of NGC 613 is the strong, multiple spiral arms emerging from
the end of a narrow, straight internal bar. Even more striking is the double nature of the major arms
emerging from either end of the bar (Fig.1). It is very tempting to identify this pair of double arms with 
a double ejection event which has taken place in the interior of the galaxy. 

The justification would be as follows: In an article in Scientific American in 1963 Arp suggested that 
ejections from galaxy interiors under the influence of rotation were the cause of spiral features. In 1964 
Lin and Shu proposed a mathematical model where spiral arms were the result of density waves in a 
disk. In 1969 Arp argued again that the empirical morphology of spirals favored an ejection origin for 
the arms. (see Arp 1986 and references therein.) Arp and Madore (1987) reported that a survey of more 
than 77,000 galaxies in the "Catalogue of Southern Peculiar Galaxies and Associations" showed that 
about 0.5 percent of peculiar galaxies have as many as three arms. This would be counter evidence 
against symmetrical density waves and at the same time positive evidence that the rare spiral arm 
configuration in NGC 613 was connected to the rare double pairs of ejected radio sources. 

It is intriguing to note that in the infrared, 7 micron image shown in NED, the inner bar is very 
narrow and straight and ends on dense knots of what are presumably intense star forming 
regions. If the two NVSS sources are real they would be associated with these two regions. 
The two inner radio sources would be associated with slight bulges of the nucleus along this 
same direction. This could be construed as outflow of ejected material at this point causing 
the star formation. In another barred spiral, NGC 1672, there are strong X-ray sources at a 
similar point which may mark the exit of ejected quasars (Arp 1998, Fig. 3-29). 

If the ejection model is correct it offers intriguing opportunities to learn about the properties of the
quasars and the spiral arms which either guide their ejection or are caused by their ejection. Since the
generally accepted optical synchrotron radiation of the quasars requires magnetized plasma it is 
suggestive that measures show magnetic fields running along spiral arms. On the observational side 
an object of z = 0.1 has been observed near a disturbance in the spiral arm of the spectacular face-on 
spiral NGC 1232 (Arp 1987). More recently two quasar like objects of z = .24 and z = .39 have been 
observed in the spiral arm-like connection between the Seyfert NGC 7603 and its +8,400 km/sec higher 
redshift companion (Lopez-Corredoira and Guitierrez 2003).  

\section{Additional Objects Possibly Associated with NGC 613}

There is a quasar 13.6' E of NGC 613 in Fig. 3  which is very bright at 15.7 apparent magnitude
and is a very strong X-ray source. Quite closely aligned on the other side of NGC 613 is a gamma ray
burster, 4b 940216, (located 17.2 arcmin W, just beyond the z = 1.69 quasar in Fig. 3).  This pair of 
unusual objects fits the distance and alignment criteria of ejected objects from bright galaxies. 
Association of GRB's with active, low redshift galaxies has been pointed out by G. Burbidge (2003) 
and Arp (2003). 

It is also notable that the bright quasar has z = .699 and only $17^"$ N of it there is a quasar of 
z = 1.177. When the latter redshift is transformed into the rest frame of the brighter object it 
becomes $z_0 = .28$ - very close to the Karlsson preferred redshift peak of $z_p = .30$. 
If physically associated it would suggest a secondary ejection.

\begin{figure}[h]
\includegraphics[width=8.5cm]{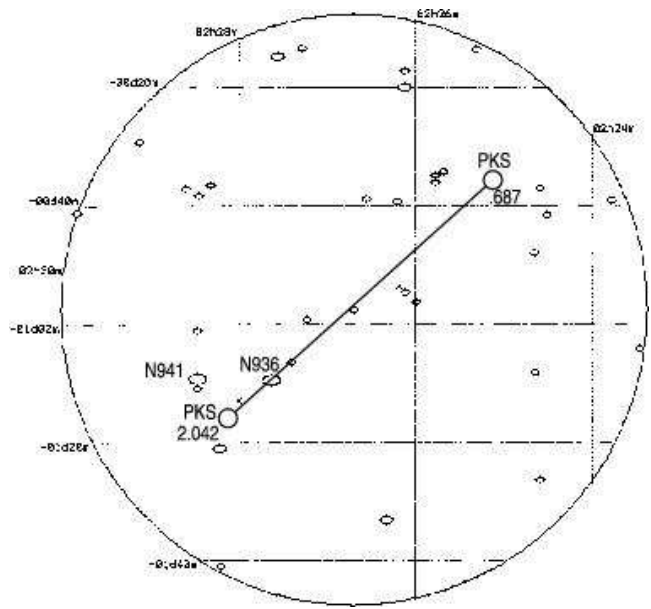}
\caption{All Simbad objects within a 50 arcmin radius are plotted. The two bright PKS quasars are
seen to be aligned across NGC 936.
\label{fig8}}
\end{figure}

\begin{figure}
\includegraphics[width=8.5cm]{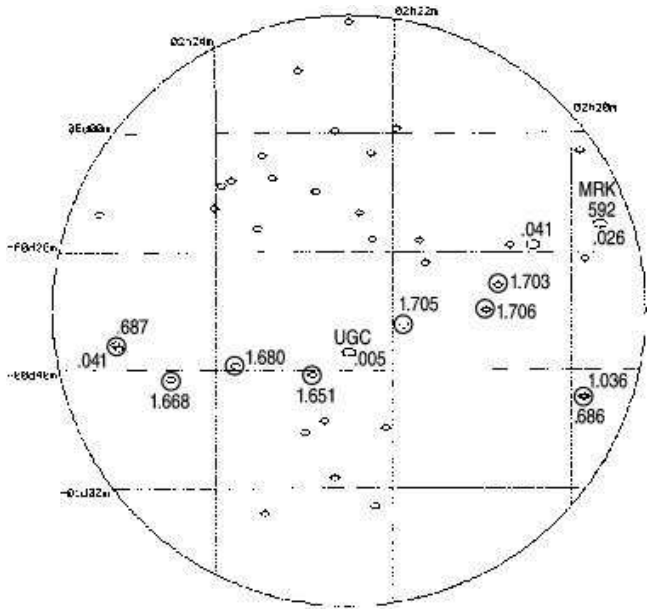}
\caption{All catalogued quasars (including SDSS) within 50 arcmin radius of a center about
32 arcmin W of PKS 0222-00. The six quasars with $1.651 \leq z \leq 1.706.$ are circled plus two
others.
\label{fig9}}
\end{figure}

\section{The more distant pair across NGC 936}

In Fig. 8 it is seen that about 50 arcmin NW of NGC 936 there is a very
bright Parks radio source (PKS 0222-00) {\it It is exactly aligned with bright PKS 
quasar SE across NGC 936.} This quasar (PKS 0222-00) is a strong X-ray source 
of 70 cts/ksec and z = .687.

Interestingly, only 1 arcmin from this quasar, which is listed as 18.4 mag., there is a 
galaxy optically elongated back toward the quasar with a faint but definite jet 
pointing in the opposite direction. The galaxy is UGC 01876 at 15.2 mag. (NED). It is 
an X-ray source (WGA) and lies within the strong nvss radio extension from 
the quasar. Perhaps most interesting of all, the high radio resolution of FIRST 
shows two compact nuclei in the quasar aligned approximately back to the 
compact radio nucleus of the galaxy. (Or possibly the radio source is a few arcsec
W of the position of the galaxy.)

If we correct the observed redshift of the quasar, $z_Q = .687$ to that of the galaxy 
$z_G = .041$  its intrinsic redshift becomes $z_0 = .620$. This is very near the 
redshift periodicity peak of $z_p = .60$. As a consequence every indication is that 
the active UGC galaxy has ejected this extremely powerful and active quasar.

We will discuss the possible relation of this quasar-galaxy pair to NG936 in the 
final section but first it is necessary to show a very unusual string of quasars in 
the vicinity of the quasar we have just discussed.

\section{Six similar redshift quasars in a string from PKS 0222-00}

Fig. 9 shows the distribution of all Cataloged quasars (including SDSS) 
within a 50 arcmin radius of a center about 32 arcmin W of PKS 0222-00. There are 
six quasars with $1.651 \leq z \leq 1.706.$ They are circled and labeled in Fig. 9. It is 
clear that these six quasars form a filament or arc leading across the field from the 
active AGN PKS 0222-00. Quasars with redshifts such as these ($<z> = 1.686$) are 
quite rare as can be seen by noting their absence in the rest of the field.

The most remarkable property of these six quasars, however, is that if we take the 
active AGN PKS 0222-00 as the origin of this string and transform the redshifts in 
the string to $z_AGN = .687$ we get six $z_0's$ with an average $z_0 = .592.$ Only 
.008 away from the Karlssson periodicity peak of $z_p = .60$!

This certainly appears as if the very powerful and active PKS AGN had ejected this 
string of quasars all of the same age (and hence the same intrinsic redshift). 
Unfortunately the origin of this string of quasars has an alternative which leaves its 
identification somewhat uncertain.

\subsection{Another possible origin}

As shown in Fig. 9 there is a galaxy of z =.005 (UGC 01839) at the center of the string 
of $z\sim1.7$ quasars. The six quasars are situated three each side, pair wise across 
the galaxy and, if ejected, with slight rotation as they become more separated from 
the galaxy. Also as shown in Fig. 8 there is a galaxy of $z = .041$ at each end of this 
string of quasars. This would speak against the origin of the string with the PKS, 
$z = .687$, AGN.

There are two arguments, however, against UGC 01839 as the origin. It is classified as a 
15.2 mag., edge on spiral with no observable nuclear bulge but does not appear active 
in radio or X-rays. The second argument is that transforming the $z\sim1.7$ quasars to 
z = .005 does not place them on a periodic redshift as an origin in the PKS object would.
As a result we prefer the PKS origin for the string but recognize that then the z = .005 
galaxy either is accidental in its position or there is more to the story than we presently 
understand. 

\section{Conclusion}

There are three major aims of this paper. One is to report radio extensions which 
suggest ejection of material along the bar in NGC 613. Since the adjacent, high redshift 
quasars are found in the same line, an ejection origin for these quasars is 
strongly supported.

The second aim is to show a completely different but confirming example of another 
barred spiral, active in both radio and X-rays, which appears to be ejecting a pair of 
quasars of similar redshift in almost exactly the same configuration as in NGC 613.

Thirdly, a strong radio connection between NGC 941 and the nearby z = 1.130 quasar is 
displayed.
  
Other connections between low redshift objects and quasars in the general region of 
NGC 936 are also discussed. I hope these cases demonstrate that more information 
on the physical nature of galaxy-quasar associations and the periodicity of the quasar 
redshifts is to be had by further analysis of the  new large area 2dF and SDSS surveys.

\medskip

Acknowledgement

I would like to thank Barry Madore who called my attention to the radio emission in the 
direction of the adjacent quasars in NGC 613, an unusually disturbed, multi-armed spiral
which we had the pleasure of discovering while producing our Catalog of Peculiar 
Southern Peculiar Galaxies and Associations.

\end{document}